# Half-metallic state and magnetic properties versus the lattice constant in $Zr_2RhZ$ (Z = Al, Ga, In) Heusler alloys

X.T. Wang, [1] J.W. Lu, [2] H. Rozale, [3] X.F. Liu, [4] Y.T. Gui, [1] G.D. Liu, [1, *]

[1] *School of Physics and Electronic Engineering, Chongqing Normal University, Chongqing 400044, PR China*

[2] *Qinggong College, North China University of Science and Technology, Tangshan 063009, PR China*

[3] *Condensed Matter and Sustainable Development Laboratory, Physics Department, University of Sidi-Bel-Abbès, 22000 Sidi-Bel-Abbès, Algeria*

[4] *Department of Materials science and Engineering, The University of Tennessee, Knoxville, TN 37996, USA*

[*] *Corresponding author. Tel./fax: +86 26564070.*
*E-mail address: gdliu1978@126.com ( prof. G.D. Liu). Address for correspondence: School of Material Sciences and Engineering, Hebei University of Technology, 8 DingZiGu 1st Road, Tianjin, PR China.*

ABSTRACT The half metallic and magnetic properties of $Zr_2RhZ$ (Z = Al, Ga, In) alloys with an $Hg_2CuTi$-type structure were systematically investigated using the first-principle calculations. $Zr_2RhZ$ (Z = Al, Ga, In) alloys are predicted to be half-metallic ferrimagnets at their equilibrium lattice constants. The $Zr_2Rh$-based alloys have $M_t$ (the total magnetic moment per unit cell) and $Z_t$ (the valence concentration) values that in agreement with Slater-Pauling rule $M_t = Z_t - 18$. The half-metallic properties and the magnetic properties at different lattice constants are discussed in detail. We expect that our results may trigger $Zr_2RhZ$ (Z = Al, Ga, In) applying in the future spintronics field.

*Keywords: Half-metallicity, Electronic structure, Heusler alloys.*

## 1. Introduction

Half-metallic (HM) materials [1] exhibiting a 100% spin polarization around the Fermi surface have received growing interest due to their interesting physical properties and potential applications in spintronic devices [1-3]. In detail, half-metals can be used as spin injectors for magnetic random access memories and other spin-dependent devices. [2]. Since the first half-metallic (HM) magnet NiMnSb, a half-Heusler alloy, was predicted by de *Groot et al.* in 1983 [1], a lot of alloys were predicted to be half-metallic materials. In fact, investigating and searching for new HM materials are mainly focusing on the Heusler alloys due to their excellent properties: high Curie temperature $T_c$, high spin polarization, low saturation of magnetization and ease of fabrication.

From the view of structural point, Heusler family can be described by two variants: full-Heusler $X_2YZ$ phases which typically crystallize in $Cu_2MnAl$ ($L2_1$) - type structure and the half-Heusler $XYZ$ variants with $C1_b$ structure. In full-Heusler alloys, if X atom is more electronegative than Y, the $Cu_2MnAl$ - type is obtained, and if the valance electron of Y atom is larger than X, the $Hg_2CuTi$ - type is obtained. Till now, many $Mn_2$, $Ti_2$, and $Sc_2$ - Heusler alloys with $Hg_2CuTi$ structure reported are HM magnetic [3-10], or even spin gapless semiconductor [11,12]. Among them, the electronic structures and magnetic properties of $Mn_2CoAl$ with $Hg_2CuTi$ structure have been theoretically investigated by us [5] in 2008. Then, in 2013, S. Ouardi *et. al.* [11] checked the crystalline structures of $Mn_2CoAl$ by X-ray diffraction (XRD) and performed the transport properties by using a physical properties measurement system

(PPMS). Predictions of $Zr_2RhZ$ (Z = Al, Ga, In) alloys with full-Heusler structures has never been made theoretically and experimentally, to the best of our knowledge. The valance electrons of Rhodium is larger than Zirconium and thus $Zr_2RhZ$ (Z = Al, Ga, In) alloys are assumed to crystallize in $Hg_2CuTi$-type structures, similar to $X_2CoZ$ (X = Sc, Ti, Mn, Zr) [5, 12-15].

Due to robust half metallic properties were reported in Zr-based Heusler-type alloys by Xie *et. al* [18], recently. A study on the half-metallic states and magnetic properties versus the lattice constant of newly designed $Zr_2$-based HM materials would be necessary. In addition, for till now, most Heusler-type HM materials are mainly composed of 3d transition metal elements. Thus, the study of Heusler alloys composed of 4d transition metal elements will enlarge the scope of exploring new functional materials. In this work, we use the first-principles method to calculate the electronic structures and magnetic properties of the $Zr_2RhZ$ (Z = Al, Ga, In). It is found that $Zr_2RhZ$ (Z = Al, Ga, In) alloys are half-metallic ferrimagnets. We also discuss the HM stability and magnetic properties under hydrostatic strain and tetragonal distortion.

## 2. Computational details

The CASTEP code was used to calculate the electronic structures, magnetism and the total energy. The CASTEP code is based on the density functional theory (DFT) plane-wave pseudo-potential method [19, 20]. In this calculation, the ultrasoft pseudo-potential [21] was used to describe the interactions between the valence electrons and the atomic core and the exchange-correlation potential was dealt with by the generalized-gradient approximation (GGA) in the scheme of Perdew–Burke–Ernzerh (PBE) [22,23]. The cut-off energy of the plane wave basis set is 450 eV for all of the cases, and a mesh of $12\times12\times12$ k-points in the full Brillouin Zone have been employed. The calculations continue to ensure good convergence until the energy deviation is less than 0.00001 eV/atom.

## 3. Results and discussions

*3.1. Electronic structures and magnetic properties at the equilibrium lattice constant*

As mentioned above, For $Zr_2RhZ$ (Z = Al, Ga, In), Zirconium is less electronegative than Rhodium, thus these materials are supposed to crystallize in $Hg_2CuTi$ structure. Namely, the X atoms (Zr atoms) occupy the A (0, 0, 0) and the B (1/4, 1/4, 1/4) sites, and Y (Rh) and Z atoms are located on C (1/2, 1/2, 1/2) and D (3/4, 3/4, 3/4), respectively. First of all, an optimization of the lattice constants were carried out for $Zr_2RhZ$ (Z = Al, Ga, In) alloys with $Hg_2CuTi$ structure. The achieved equilibrium lattice constants are shown in Table 1, i.e., 6.66 Å, 6.64 Å and 6.81 Å, respectively.

Table 2 shows the calculated molecular and atomic magnetic moments for the $Zr_2RhAl$, $Zr_2RhGa$, and $Zr_2RhIn$ alloys under their equilibrium lattice constant. From table 2, we can see that $Zr_2RhAl$, $Zr_2RhGa$, and $Zr_2RhIn$ have integer magnetic moments of 2 $\mu_B$. As we know, the integer value of the total magnetic moment is characteristic of half-metallic materials. Moreover, The atomic magnetic moment of Zr (A)/Zr (B) are 1.60$\mu_B$ / 0.6$\mu_B$, 1.54$\mu_B$ / 0.72$\mu_B$, 1.54$\mu_B$ / 0.78$\mu_B$ and the atomic magnetic moments of Rh(C)/Z are -0.1$\mu_B$ / -0.1$\mu_B$, -0.1$\mu_B$ / -0.18$\mu_B$, -0.16 $\mu_B$ / -0.18$\mu_B$, respectively. We can see that the major contributions to the total magnetic moments come from Zr(A) and Zr(B) atoms, while the Rh(C) and Z atoms can be regards as minor. The atomic magnetic moments of Zr(A) and Zr(B) are different from each other, indicating different atomic environment. And the atomic magnetic moment of Rh/Z is antiparallel to that of Zr(A)/Zr(B) atom, Thus, Zr$_2$RhAl, Zr$_2$RhGa, and Zr$_2$RhIn alloys are HM ferrimagnets.

When we consider Zr$_2$RhAl, Zr$_2$RhGa, and Zr$_2$RhIn alloys to be Heusler-type, the number of the valence electrons are 20. Our calculated results show that the total spin magnetic moments are integral values of 2 $\mu_B$. These results indicate that the total magnetic moment $M_t$ of Zr$_2$RhAl, Zr$_2$RhGa, and Zr$_2$RhIn alloys has a linear relationship to the number of valence electrons $Z_t$: *$M_t = Z_t$-18,* which is in consistent with Zr-based Heusler-type HM materials as mentioned above [13, 17].

Table 1 shows the conduction band minimum (CBM), the valence band maximum (VBM), and the size of the gap of the spin-down of Zr$_2$RhZ (Z = Al, Ga, In) at their equilibrium lattice constant. For Zr$_2$RhAl/Ga/In, the VBM is -0.2839/-0.3829/-0.3062 eV, and the CBM is 0.2430/0.2618/0.3517 eV, turning out a 0.3517/0.3517/0.6579 eV band gap. Further, the spin-flip gaps (HM gap), which is the minimum energy required to flip a minority-spin electron from the valence band maximum edge to the majority-spin Fermi level, are the most proper indication of the half-metallicity of a material. The HM gaps of $Zr_2RhZ$ (Z = Al, Ga, In) alloys are 0.2839 eV, 0.3829 eV and 0.3062 eV, respectively. This value is quite larger than the most of Zr-based Heusler-type HM materials (eg. ZrCoFeSi (0.22 eV), ZrMnVSi (0.14 eV) and ZrMnVGe (0.18 eV) ), while smaller than in $Ti_2CoSn$ (0.59 eV). The wider HM gap has a stronger ability to resist destruction of temperature stability of half-metallicity.

The electronic band structures of the $Zr_2RhZ$ (Z = Al, Ga, In) alloys along the main symmetry in the irreducible Brillouin zone are shown in Fig. 1. For these three alloys, we can see that the Fermi levels are located at a band gap in the spin down direction and have an intersection with the valence bands in the spin up direction, which indicates that $Zr_2RhZ$ (Z = Al, Ga, In) alloys are half metals and leads to 100% spin-polarization around the Fermi level.

Fig. 2 presents the spin-polarized total densities of states (DOS) and atom-projected DOS of $Zr_2RhZ$ (Z = Al, Ga, In) alloys at their equilibrium lattice constants. It is obvious that the *p* states of Z atom mainly occupied the lowest part of the valence states in both the spin up and spin down directions. The 4d electrons of Zr and Rh atoms determine the DOS around the Fermi level. In spin down channel, the DOS located above the Fermi level mainly comes from Zr(B) atoms, while below the Fermi level mainly arises from Rh(C) atoms. Similar to the case of $Zr_2CoAl$ and $Zr_2CoSn$ [13,17], the band gap in the spin channel behavior semiconducting origin from the hybridization effects of transition metals' s d orbitals.

Noted that not all of Heusler alloys can be synthesized and form stable phase. To examine that the Hg$_2$CuTi-type Zr$_2$RhZ (Z = Al, Ga, In) alloys can be synthesized in the experiment and the phase stability, the formation energy E$_{formation}$ was calculated by subtracting the sum of equilibrium total energies for constituent elements (Zr and Rh with HCP structures, Al, Ga, In with FCC structures), from the equilibrium total energies of corresponding alloys under current study, using the formula:

$E_{formation} = (E_{tol} - 2 E_{Zr} - E_{Rh} - E_Z )$, we find that the E$_{formation}$ of these three alloys are -2.18 eV, -2.22 eV, and -2.42 eV, respectively. The calculated formation energy turns out to be negative and comparable for the same kind of compounds [13, 14]. Negative formation energy means the compound is energetically stable. Hence it is possible to be synthesized experimentally.

*3.2. Effect of hydrostatic strain on the electronic structures and magnetic properties*

During the process of non-equilibrium melt-spun or ball milling, the hydrostatic strain has uncontrolled change, which may make the lattice constant of these HM materials to change and deviate from the ideal one, and affect their half-metallicity. Moreover, when Zr$_2$RhZ (Z = Al, Ga, In) are synthesized by certain non-equilibrium preparation methods, for example, molecular beam epitaxy, the hydrostatic strain induced by the substrate also can not be ignored. Thus, it is necessary to study the electronic structures and magnetism versus the lattice constant in Zr$_2$RhZ (Z = Al, Ga, In) alloys for the future practical applications.

To discuss the stability of half metallic states versus the lattice constant, the electronic structures at different lattice constants were calculated for these three alloys. The electronic structures are quite similar under different lattice constants, thus, the CBM and VBM in the spin down channel have been used to represent the half metallic properties. The negative sign means that the energy of VBM or CBM are lower than the Fermi level, and the positive sign means the energy of VBM or CBM are higher than Fermi level. Fig. 3 shows the CBM, VBM, the position of the Fermi level in spin down channel and the width of the band gap as a function of the lattice constant. We can see that a stable band gap can be kept in spin down channel for a large lattice constant range. Thus, these alloys are suitable to be an ideal candidate for spintronics.

When the lattice constants are compressed to the value of 6.327 Å / 6.308 Å / 6.333 Å, the Fermi level falls within the conduction band. That is to say, the half metallic properties have been broken for the $Zr_2RhAl$, $Zr_2RhGa$, and $Zr_2RhIn$ alloys under a hydrostatic contraction strain more than 5% / 4% / 7%, so that $Zr_2RhAl$, $Zr_2RhGa$, and $Zr_2RhIn$ alloys have already became conventional ferrimagnetisam without HM behavior and the spin polarization decreases.

In order to further observe the magnetic moment changing trend with varying lattice constant, the total and atomic magnetic moment as a function of the lattice constant for the $Zr_2RhAl$, $Zr_2RhGa$, and $Zr_2RhIn$ alloys are shown in Figs. 4. From Fig. 4, we can see that the values of total magnetic moment for the $Zr_2RhAl$, $Zr_2RhGa$, and $Zr_2RhIn$ alloys are still $2\mu_B$ in the range of 6.372-7.259, 6.308-7.238, and 6.333-7.422, respectively. Meanwhile, the total magnetic moment still follows the SP rule $M_t = Z_t-18$. And the Zr (B) and Rh atoms are sensitive to the lattice constant, while the Zr (A) and Z atoms are not. For $Zr_2RhZ$ (Z = Al, Ga, In) alloys, the magnetic moments of Zr(B) atoms increase when the lattice constant expands. And with the lattice constant expands, the values of Rh atoms keep gradually decreasing.

*3.3. Effect of tetragonal deformation on the electronic structures and magnetic properties*

As we known, the growth of thin-film materials is an important technique in practical applications in which tetragonal deformation is most likely to occur due to films tends to adjust their in-plane lattice constants to the substrate while changing the out-of –plane lattice constant to keep the volume of the unit-cell almost the same as the equilibrium bulk lattice constant. For simulating the case of tetragonal deformation, we fix the unit-cell volume in the equilibrium bulk volume (a×a×a = $a^3$), then change the c/a ratio.

The calculated CBM, VBM, and the total and atomic magnetic moment as a function of the c/a ratio in the range of 0.91 – 1.09 for $Zr_2RhZ$ (Z = Al, Ga, In) alloys are shown in Fig. 5 and Fig. 6, respectively. As shown in Fig. 5, $Zr_2RhZ$ (Z = Al, Ga, In) can maintain their HM states when c/a ratio are changed by 0.92 – 1.09, 0.91 – 1.09, 0.91 – 1.09, respectively. It appears that the HM character of $Zr_2RhZ$ (Z = Al, Ga, In) alloys exhibits a low sensitivity to the tetragonal deformation. As the case of $Zr_2RhAl$ when c/a = 0.91, the conduction bands in the spin down channel have an overlap with the Fermi level, and the half metallic gap is destroyed. For these three alloys, the band gap and the HM gap decrease when c/a ratio ranges from 1.00-1.09 and 1.00-0.91.

Next, we focus on the magnetic properties, as shown in Fig. 6. Either the total or atomic magnetic moment exhibits a low sensitivity to the tetragonal deformation. The value of the total or atomic magnetic moment hold nearly unchanged in the whole range expect the case of c/a = 0.91. For this case, $Zr_2RhAl$ alloys have already became conventional ferrimagnetisam and the total magnetic moment is 1.95$\mu_B$.

## 4. Conclusions

First principles investigations of half-metallic state and magnetic properties versus the lattice constant in $Zr_2RhZ$ (Z = Al, Ga, In) Heusler alloys have been reported during our current work. It has been found that $Zr_2RhZ$ (Z = Al, Ga, In) alloys are HM ferrimagnets. The total magnetic moments of Hg2CuTi-type $Zr_2RhZ$ (Z = Al, Ga, In) are 2$\mu_B$ per unit cell, which is following the SP rule $M_t = Z_t - 18$. For $Zr_2RhZ$ (Z = Al, Ga, In) alloys, their half-metallicity is robust against hydrostatic strain and tetragonal deformation, making these alloys very stable with respect to the polarization properties. Noted that these alloys have negative formation energies, so that these alloys are possible to be prepared as multi-layer by the method of molecular beam epitaxy or others.

**Acknowledgments**

This work was supported by the Hebei Province Higher Education Science and Technology Research Foundation for Youth Scholars (No. Q2012008), Hebei Province Natural Science Foundation (No. E2013202181)，The Basic and Frontier Research Project of Chongqing City (No. cstc2013jjB50001). One of the authors (G.D. Liu) acknowledges the financial support from Hebei Province Program for Top Young Talents.

**Figure captions**

Fig. 1 Band structures for Zr$_2$RhZ (Z = Al, Ga, In) alloys with an Hg$_2$CuTi-type structures at their equilibrium lattice constants.

Fig. 2 Calculated total and atom-projected DOS for Zr$_2$RhZ (Z = Al, Ga, In) alloys with an Hg$_2$CuTi-type structures at their equilibrium lattice constants.

Fig. 3 CBM and VBM of the spin down band as a function of lattice constant for Zr$_2$RhZ (Z = Al, Ga, In) alloys.

Fig. 4 The total and site-projected magnetic moments as a function of lattice constant for Zr$_2$RhZ (Z = Al, Ga, In) alloys.

Fig. 5 CBM and VBM of the spin down band as a function of c/a ratio for Zr$_2$RhZ (Z = Al, Ga, In) alloys.

Fig. 6 The total and site-projected magnetic moments as a function of c/a ratio for Zr$_2$RhZ (Z = Al, Ga, In) alloys.

Table 1

The maximum of the valence band (VBM), the minimum of the conduction band (CBM), size of the gap (Gap) and the half-metal gap (HM Gap), the formation energy ($E_f$) of $Zr_2RhZ$ (Z = Al, Ga, In) alloys at their equilibrium lattice constants.

| $Zr_2RhZ$ | a(Å) | VBM(eV) | CBM(eV) | Gap(eV) | HM Gap(eV) | $E_f$ (eV) |
|---|---|---|---|---|---|---|
| $Zr_2RhAl$ | 6.66 | −0.2839 | 0.2430 | 0.5269 | 0.2839 | -2.18 |
| $Zr_2RhGa$ | 6.64 | −0.3829 | 0.2618 | 0.6437 | 0.3829 | -2.22 |
| $Zr_2RhIn$ | 6.81 | −0.3062 | 0.3517 | 0.6579 | 0.3062 | -2.42 |

Table 2

The calculated molecular and atomic magnetic moments (M) for the $Zr_2RhZ$ (Z = Al, Ga, In) alloys

| $Zr_2RhZ$ | $M_{total}$ ($\mu_B$) | $M_{Zr(A)}$ ($\mu_B$) | $M_{Zr(B)}$ ($\mu_B$) | $M_{Rh}$ ($\mu_B$) | $M_Z$ ($\mu_B$) |
|---|---|---|---|---|---|
| $Zr_2RhAl$ | 2 | 1.60 | 0.6 | -0.1 | -0.1 |
| $Zr_2RhGa$ | 2 | 1.54 | 0.72 | -0.10 | -0.18 |
| $Zr_2RhIn$ | 2 | 1.54 | 0.78 | -0.16 | -0.18 |

Fig. 1

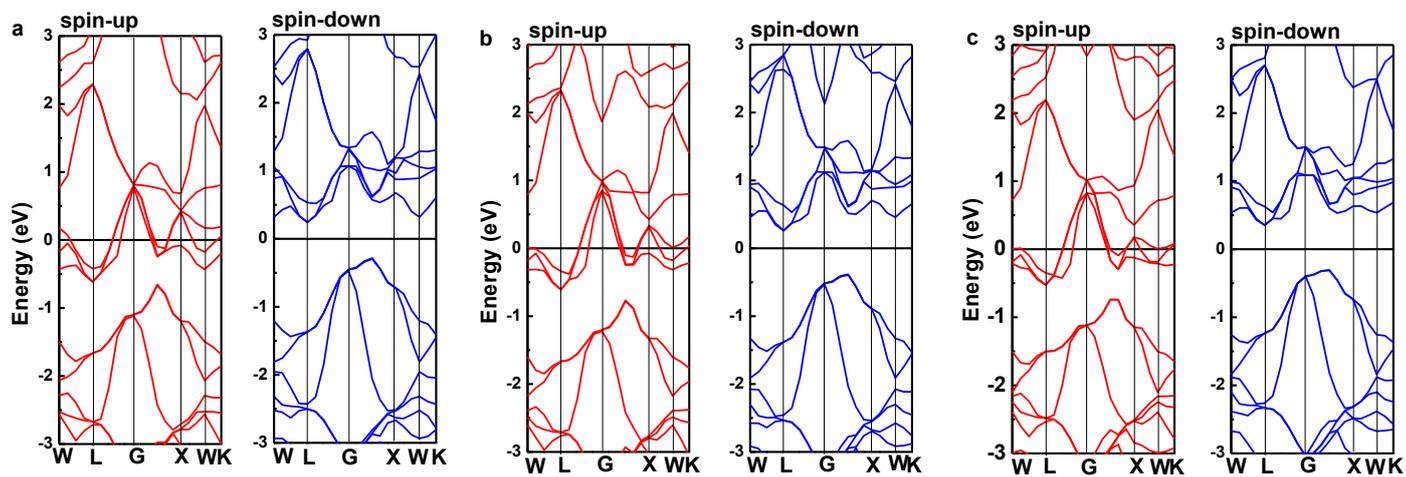

Fig. 2

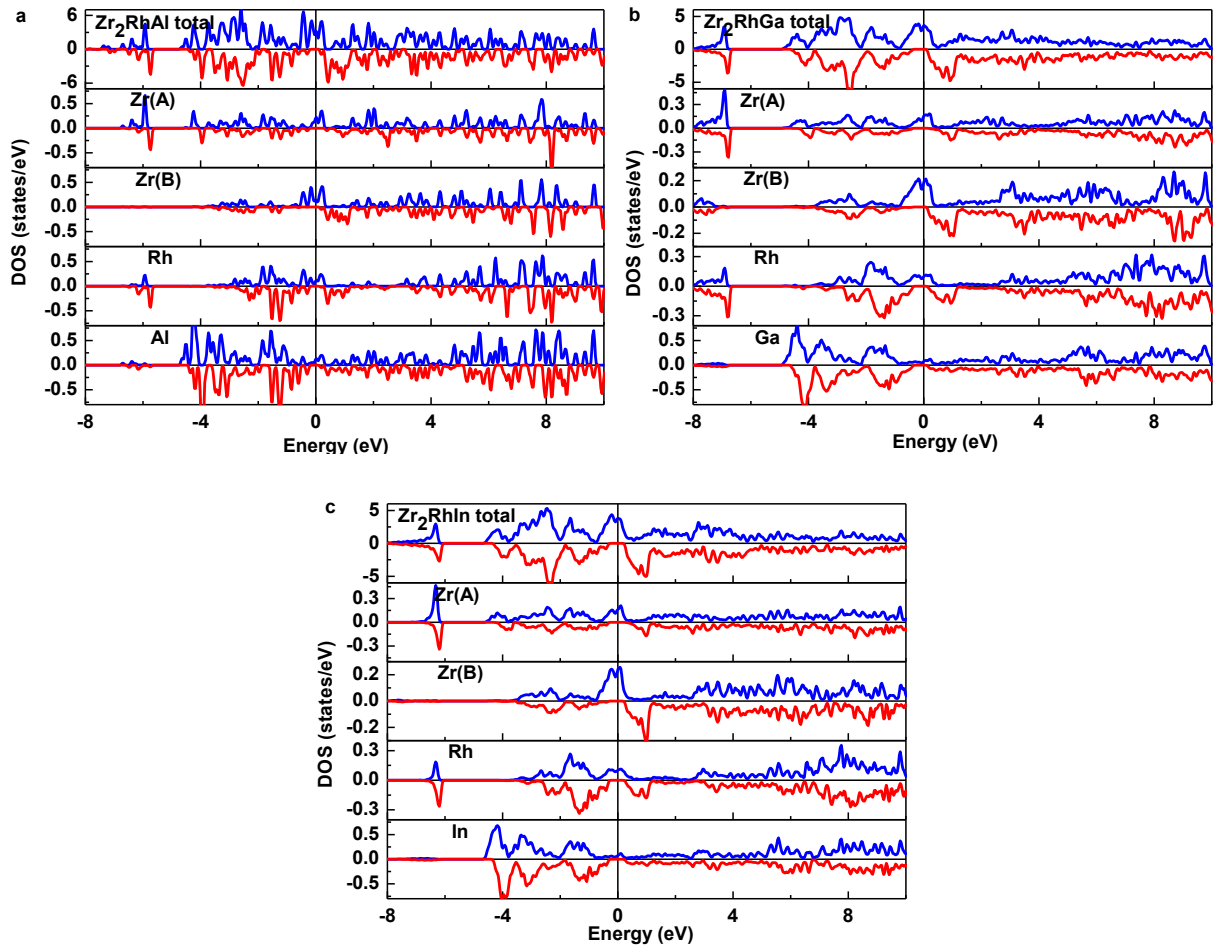

Fig. 3

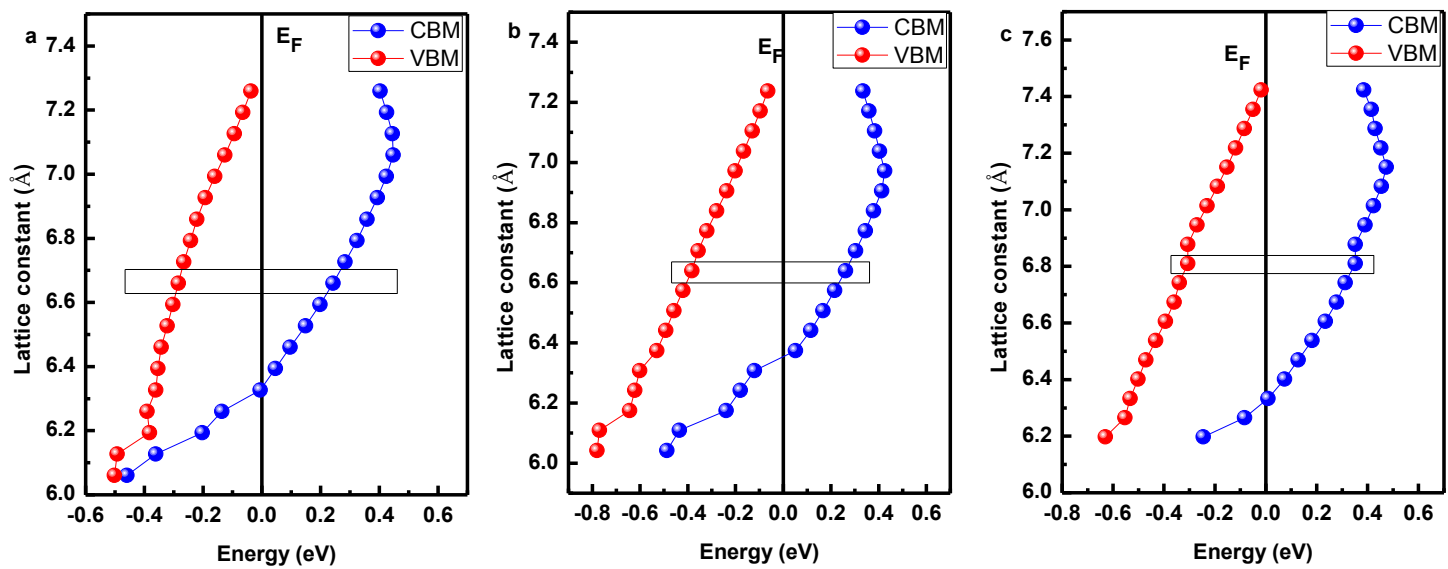

Fig. 4

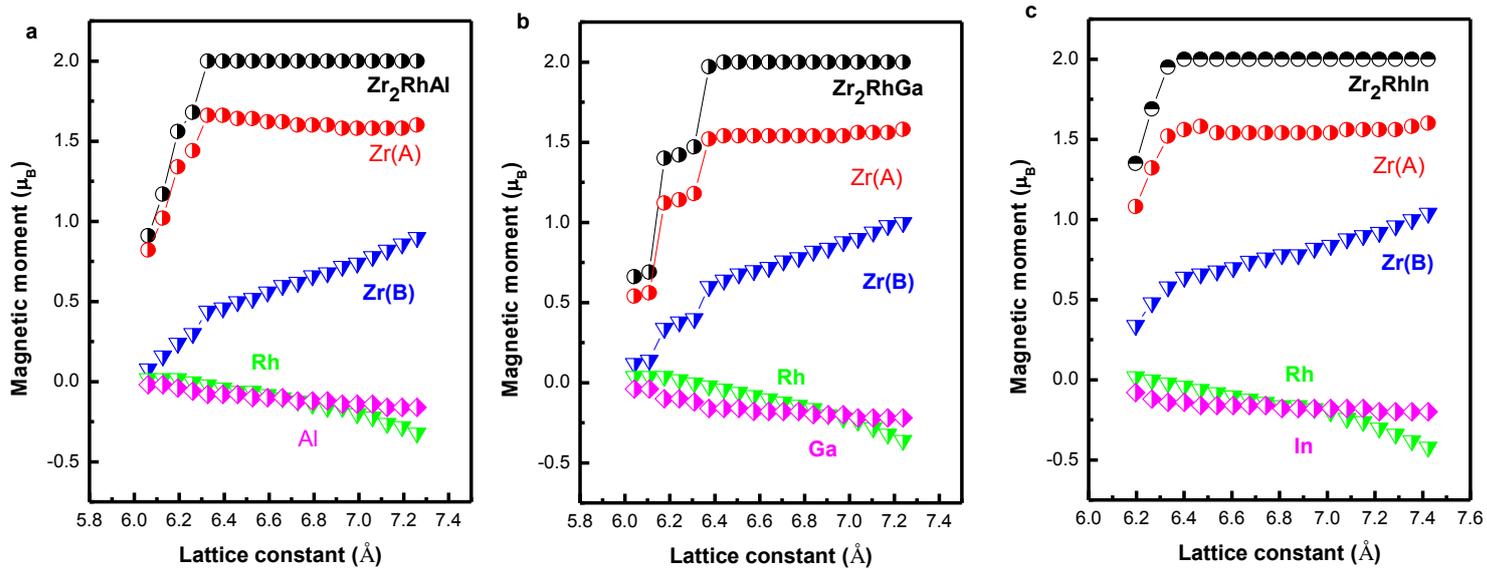

Fig. 5

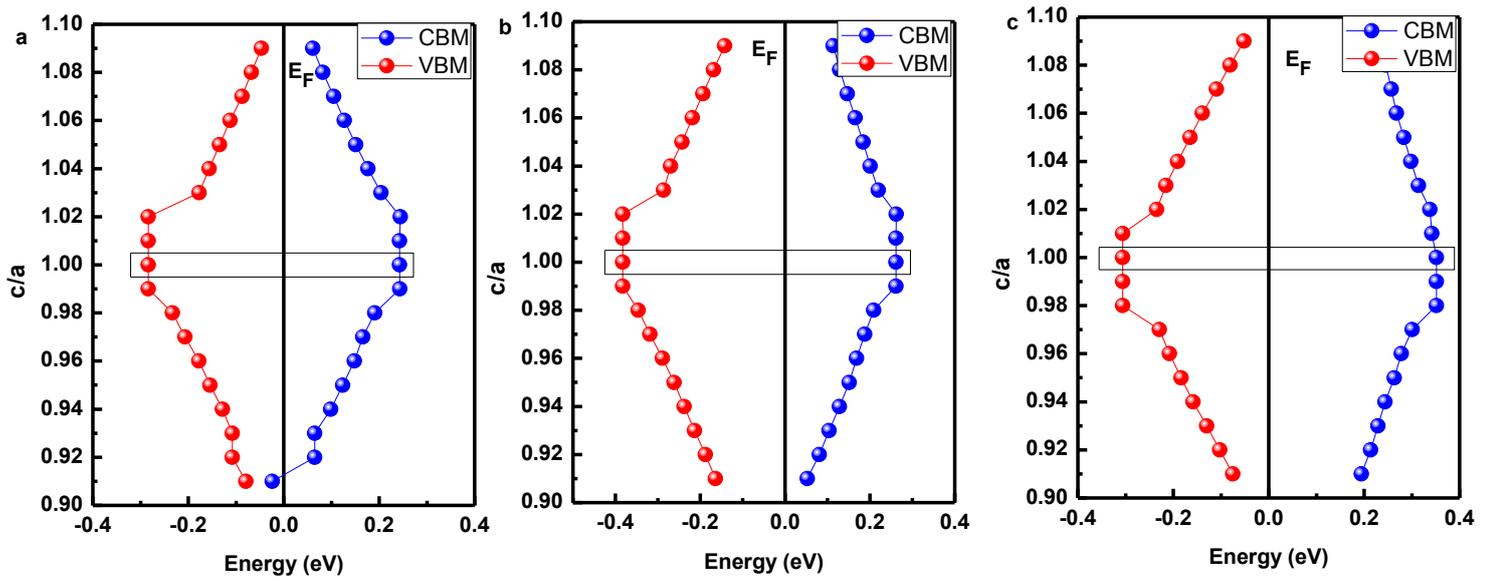

Fig. 6

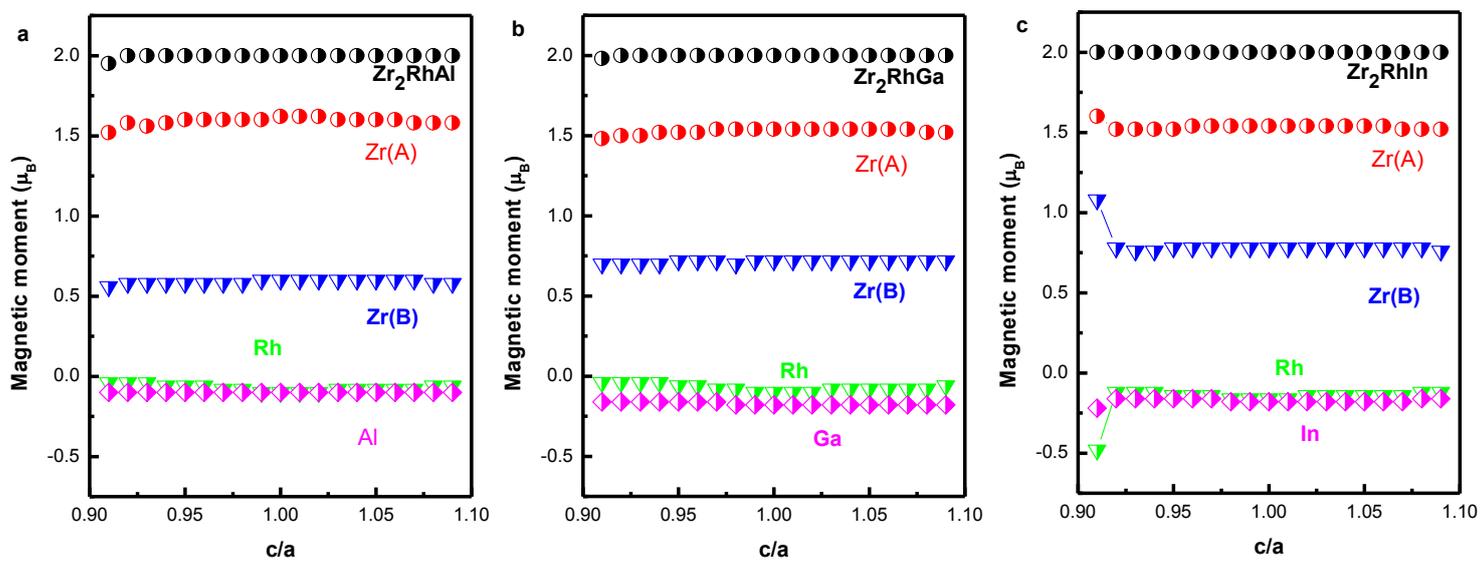